\documentclass[conference,a4paper]{IEEEtran}
\IEEEoverridecommandlockouts
\usepackage{cite}
\usepackage{amsmath,amssymb,amsfonts}
\usepackage{algorithmic}
\usepackage{graphicx}
\usepackage{textcomp}
\usepackage{xcolor}
\def\BibTeX{{\rm B\kern-.05em{\sc i\kern-.025em b}\kern-.08em
    T\kern-.1667em\lower.7ex\hbox{E}\kern-.125emX}}
\usepackage{suffix,mathtools}
\DeclarePairedDelimiterX\MeijerM[3]{\lparen}{\rparen}%
{#3\,\delimsize\vert\begin{smallmatrix}#1 \\ #2\end{smallmatrix}}
\newcommand\MeijerG[8][]{%
  G^{\,#2,#3}_{#4,#5}\MeijerM[#1]{#6}{#7}{#8}}

\WithSuffix\newcommand\MeijerG*[7]{%
  G^{\,#1,#2}_{#3,#4}\MeijerM*{#5}{#6}{#7}}

\def\BibTeX{{\rm B\kern-.05em{\sc i\kern-.025em b}\kern-.08em
    T\kern-.1667em\lower.7ex\hbox{E}\kern-.125emX}}

\makeatletter
\newcommand{\linebreakand}{%
  \end{@IEEEauthorhalign}
  \hfill\mbox{}\par
  \mbox{}\hfill\begin{@IEEEauthorhalign}
}
\makeatother

\begin{document}

\title{Performance Analysis of UAV-Assisted RF-UOWC Systems \\

\thanks{This work has received funding from the Horizon 2020 research and innovation staff exchange grant agreement No 101086387, and from the  Science Fund of the Republic of Serbia, grant number 6707, REmote WAter quality monitoRing anD IntelliGence – REWARDING. This work has been supported by   the Secretariat for Higher Education and Scientific Research of the Autonomous Province of Vojvodina through the project “Visible light technologies for indoor sensing, localization and communication in smart buildings” (142-451-2686/2021). This publication was based upon work from COST Action NEWFOCUS CA19111, supported by COST (European Cooperation in Science and Technology).}
}
\author{\IEEEauthorblockN{Tijana Devaja,\IEEEauthorrefmark{1} Milica Petkovic,\IEEEauthorrefmark{1} Marko Beko,\IEEEauthorrefmark{2} Dejan Vukobratovic,\IEEEauthorrefmark{1} 
\IEEEauthorblockA{\IEEEauthorrefmark{1}University of Novi Sad, 
Novi Sad, Serbia}
\IEEEauthorblockA{\IEEEauthorrefmark{2}Copelabs, Universidade Lusofona de Humanidades e Tecnologias, Lisbon, Portugal
}}}

\maketitle

\begin{abstract}
This paper introduces a relay-assisted solution for downlink communications in a mixed system of  Radio Frequency (RF) and Underwater Optical Wireless Communications (UOWC) technologies. During the initial downlink phase, data transmission occurs via RF link between hovering Unmanned Aerial Vehicle (UAV) and the floating buoys at the water surface. As fixed buoy acts as amplify-and-forward relays, the second  UOWC link represents downlink signal transmission from the floating buoy to the underwater device.  Best relay selection is adopted, meaning that only the buoy  with the best estimated  RF-based UAV-buoy channel will perform  signal transmission to an underwater device.
Analytical expression for  the outage probability  is derived  and utilized to examine the system's performance behavior for various  UOWC and RF channel conditions.
\end{abstract}


\section{Introduction}

The number of Internet of Things (IoT) connections is forecasted to rise from 15.7 billion in 2023 to 38.9 billion by 2029 \cite{ericsson} which leads to increase of data traffic by  160$\%$ every year, and is 
expected to increase further in future \cite{fuji}. 5G and beyond-5G serves as the ideal enabler for the IoT due to its high data speed, minimal latency, enhanced mobility, low energy consumption, cost-effectiveness, and capacity to manage significantly larger numbers of devices. Over recent decades, IoT systems have been extensively researched, designed, and developed, finding widespread applications across diverse domains including industry \cite{r3a}, smart transportation \cite{ST}, smart energy \cite{SE}, smart cities \cite{r1}, healthcare \cite{r4a}, and more.

As human activities expand into the underwater environment, which covers around 70$\%$ of the Earth’s surface, there is a growing focus on underwater wireless communication (UWC). UWC technology facilitates data transmission in aquatic environments through radio-frequency (RF) waves, acoustic waves, or optical waves, enabling underwater exploration \cite{UWC2}. Underwater  optical wireless communication (UOWC) offer advantages such as higher data transmission rates over short distances, lower power consumption, and increased security \cite{UOWC1, UWC2}.
Compared to RF and
acoustic methods,  UOWC provides higher data rates for short and moderate transmission distances. While UOWC offers several benefits, it does have limitations like restricted range and dependence on water turbidity and attenuation. These factors can diminish signal quality, particularly over long distances or in murky water conditions. Nevertheless, ongoing research and advancements in optical communication technologies are steadily enhancing the effectiveness of UOWC systems \cite{acustic, parameters}. As a result, they are becoming a more practical choice for underwater communication needs and hold potential as a valuable component in the evolution toward beyond-5G wireless networks \cite{beyond5G}.

Low-cost and highly mobile unmanned aerial vehicles
(UAVs) offer versatility, speed, and cost-effectiveness in various applications.
Some of the examples where UAVs made  impact include
aerial surveillance, 
search and rescue operations,
disaster response, 
infrastructure inspection, agriculture,
environmental monitoring. UAVs are increasingly being employed for remote sensing and relaying systems due to their rapid deployment capabilities and ability to establish line-of-sight (LoS) connections without the need for complex infrastructure \cite{UAV1,UAV2a,UAV2, UAV3, UAV3a}. 
Lately, UAV-based mixed RF-UOWC systems have emerged as a promising solution for establishing connections between underwater devices and communication networks using hovering UAVs \cite{UAV4,UAV5, UAV6}.
These systems integrate both UOWC and RF communications, leveraging the versatility and mobility of UAVs to bridge the gap between traditional communication networks and  underwater environments.
 
Motivated by the insights discussed earlier, this paper aims to design and assess a novel UAV-based mixed RF-UOWC system. The focus lies on the  integration of both communication technologies, in order to optimize the performance of both communication links. 
More precisly, we analyse dual-hop amplify-and-forward (AF) relaying RF-UOWC systems. Considered scenario include  hovering UAV, a number of floating buoys at the water surface which acts as relays, and a underwater device. Considering best relay selection approach, after channel state estimation, the UAV selects the buoy with the best estimated channel state
information (CSI) to perform downlink transmission to the corresponding buoy.  The buoy  acts as the AF relay and further transmit the information to the underwater device. 
We assume that the RF link is affected by Rayleigh fading \cite{RFintro}, whereas the UOWC link encounters mixed exponential–generalized gamma (EGG) distribution fading model,
proposed in \cite{parameters}, which is used to characterize turbulence-induced
fading while considering presence of air
bubbles and temperature gradient for fresh and salty water. In presented analysis, we utilize a combined model that incorporates the mixture of EGG distributed turbulence and pointing errors \cite{model,ref8},  to characterize the UOWC environment. Following described scenario, we derive novel analytical expressions for  outage probability, which is further used to presented numerical results and have concluding remarks related to system performance behavior. 

The rest of the text is organized as follows. Section II defines the system model, including both RF and UOWC channel models, while  Section III  provides the outage probability performance analysis. Numerical results are given in Section IV and Section V concludes the paper.

\section{System model}
\begin{figure}[t!]
\centerline{\includegraphics[width=3in, height=2in]{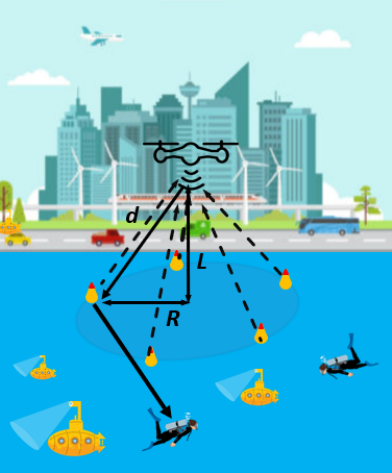}}
\caption{Considered scenario.}
\label{Fig1}
\end{figure}

In this paper we consider the dual-hop  UAV-assisted mixed RF-UOWC system, illustrated in Fig~\ref{Fig1}. The overall downlink communication involves two stages. Initially, in the downlink RF scenario signals are considered to be sent from hovering UAV, positioned at a standard distance $L$ above the water surface, to total  $N$  floating buoys, serving as AF relays. The UAV is considered equidistant from the buoys which are evenly distributed along the perimeter of a circle with a radius of $R$. Best relay selection  scheme is adopted, thus the    channel states  are estimated by sending the pilot signal from all buoys to the UAV. Following best estimated CSI, the UAV selects the best buoy to perform transmission. The second stage involves downlink UWOC scenario, where information is sent from the best buoy  to the underwater  user. 

\subsection{RF-based UAV-buoy link}

The first link illustrates the RF downlink connection between the  hovering UAV and total $N$ floating buoys. The UAV monitors the RF Rayleigh fading channel conditions through local feedback from the relays.  Best relay selection is adopted, thus the UAV performs transmission to a buoy with best estimated CSI. 
Received  signal at the buoy can be is defined as
\begin{equation}
\begin{split}
y_1=h_1 x_1+n_1
\end{split}
\label{eq1},
\end{equation}
where RF signal transmitted from the UAV to the best buoy is denoted as $x_1$  with an average
power $P_1$, and the fading amplitude is denoted as $h_1$. The additive white Gaussian noise (AWGN) in RF link,
with zero mean and variance $\sigma_1^2$, is denoted as $n_1$. The instantaneous signal-to-noise Ratio (SNR) over the RF link is given by
$\gamma_1 = \frac{h_1^2 P_1 g_1}{\sigma_1^2} = h_1^2\mu_1, 
\label{snrRF}$
where the average SNR is defined as
$\mu_1 = \frac{P_1 g_1}{\sigma_1^2}. 
\label{AVsnrRF}$
The average power gain of the RF link, denoted by $g_1$, is defined as \cite{sensors}
 \begin{equation}
g_1 = G_0 d^{-2} = \frac{G_0}{R^2 +L^2} 
\label{gain},
\end{equation}
where $d$  represents distance between each buoy and the UAV ($d^2 = R^2 + L^2$), and $G_0$ denotes the channel power gain at the reference distance $R_0=1$ m.

Further, after signal is received at the best buoy, it is amplified by a  relay gain $G$. As fixed gain AF relay is adopted, the amplification is performed based on the statistical CSI, leading to the  $G$ definition \cite[eq. (3)]{sosa}
\begin{equation}
\begin{split}
G = \frac{P_{\rm r}}{\mathbb{E} [h_1^2]P_1g_1 + \sigma_1^2 } = \frac{P_{\rm r}}{\sigma_1^2 }  \frac{1}{\mathbb{E} [\gamma_1] + 1 } = \frac{P_{\rm r}}{\sigma_1^2 C}  
\end{split}
\label{gainAF},
\end{equation}
where  $P_{\rm r}$ is the relay output signal power, $C=\mathbb{E} [\gamma_1] + 1 $ represents a constant determined by the relay gain, and $\mathbb{E} [\cdot]$ denotes mathematical  expectation.

Considering the best rely selection and that RF link experiences Rayleigh fading,  the probability density function (PDF) of the instantaneous SNR in the first hop, $\gamma_{1}$, is  determined as \cite[eq. (9) for $k=N$]{feedback}
\begin{equation}
\begin{split}
f_{\gamma_{1}}(x)=N\sum_{k=0}^{N-1}\binom{N-1}{k}\frac{(-1)^k}{\mu_1}e^{-\frac{(k+1)x}{\mu_1}}
\end{split}
\label{pdf},
\end{equation}
while the cumulative distribution function (CDF) of $\gamma_{1}$ is defined as
\begin{equation}
\begin{split}
F_{\gamma_{1}}(x)=1-N\sum_{k=0}^{N-1}\binom{N-1}{k}\frac{(-1)^k}{(k+1)}e^{-\frac{(k+1)x}{\mu_1}}
\end{split}
\label{cdf}.
\end{equation}
The constant  determined by the relay gain can be determined based on \eqref{pdf} as \cite[eq. (6)]{sosa}
\begin{equation}
\begin{split}
C = 1 +\!N \sum_{k=0}^{N-1}\binom{N-1}{k} \frac{(-1)^k \mu_1}{(k+1)^2}
\end{split}
\label{constant}.
\end{equation}

\subsection{UOWC-based buoy-user  link}

The second stage corresponds to a downlink UOWC transmission. In this phase, the best among  $N$  relays transmit amplified signal to the underwater user. The floating buoys positioned on the water surface, are equipped with OWC transmitters, such as LEDs or laser diodes, utilizing intensity modulation with on-off keying. Underwater device is equipped with OWC receiver, i.e.,  photodetector, which facilitate direct detection and conversion of optical signals to electrical signals.
Received electrical signal at the underwater user device is defined as 
\begin{equation}
\begin{split}
y_2= G P_2 \eta I x_2 +n_2
\end{split}
\label{eq1a},
\end{equation}
where $P_2$ denotes the transmitted optical power,  $x_2$ stands for the signal transmitted by the active user,  $I$ signifies the normalized irradiance resulting from underwater turbulence and pointing errors, and $\eta$ represents the optical-to-electrical conversion coefficient. The additive noise, denoted by $n_2$, is modeled as a zero-mean Gaussian random variable with variance $N_0$.

The instantaneous SNR over the  UOWC link is defined   as \cite{ref8},
$\gamma_2 = \frac{G^2 P_2^2 \eta^2 I^2}{\mathbb{E}[I]^2} = \frac{I^2}{\mathbb{E}[I]^2}\mu_2$,
 where the  average electrical
SNR is defined as 
$\mu_2 = \frac{ G^2 P_2^2 \eta^2 \mathbb{E}[I]^2 }{N_0}$.
It is further  related to the average SNR as
$\Bar{\gamma_2}= \frac{ \mu_2 \mathbb{E}[I^2]}{\mathbb{E}[I]^2}$.
The UOWC link experiences influences from both underwater optical turbulence, which is described by a mixture  EGG distribution, and pointing errors, as indicated in references \cite{model, ref8}. The $n$-th moment of $I$ is derived as \cite{ref8} 
\begin{equation}
\begin{split}
\mathbb{E}[I^n]=w\frac{(\lambda A_0)^n n! \xi^2}{n+\xi^2}+(1-w)\frac{(b A_0)^n \Gamma(a+\frac{n}{c})\xi^2}{\Gamma(a)(n+\xi^2)}
\end{split}.
\label{gama1nad1}
\end{equation}
Parameters $w$, $\lambda$, $a$, $b$, and $c$ describe the turbulence properties associated with the mixture EGG distribution, which vary based on factors like water salinity and bubble level (BL). These parameter values are specified in \cite{parameters} and shown in Table I.
The parameters $A_0$ and $\xi$ are linked to pointing errors.

The PDF of the instantaneous SNR $\gamma_2$, influenced by both the mixture  EGG fading turbulence and pointing errors, is  given in \cite[eq. (3)]{model}
\begin{equation}
\begin{split}
 f_{\gamma_2}(x)  \! &= \! \frac{ w \xi^2}{x}  \MeijerG*{2}{0}{1}{2}{\xi^2+1}{1, \xi^2}{ \frac{x}{\lambda A_0\rho}}  \\
& + \frac{ (1-w)\xi^2 } {\Gamma(a) x } \MeijerG*{2}{0}{1}{2}{\frac{\xi^2}{c}+1}{a,\frac{\xi^2}{c}}{ \frac{x^c}{(b A_0 \rho)^c}}
\end{split}
\label{UOWC_pdf},
\end{equation}
where  $\rho=\Bar{\gamma_2}/\mathbb{E}[I]^2$, $\MeijerG*{m}{n}{p}{q}{\cdot} {\cdot}{\cdot}$ defines the  Meijer’s \textit{G}-function \cite[(9.301)]{grad} and  $\Gamma(z)$ is the Gamma function \cite[(8.31)]{grad}. 
According to \cite[eq. (4)]{model}, the CDF  of $\gamma_2$ is defined as 
\begin{equation}
\begin{split}
 F_{\gamma_2}(x)  \! &= w\xi^2\MeijerG*{2}{1}{2}{3}{1, \xi^2+1}{1, \xi^2, 0}{ \frac{x}{\lambda A_0\rho}} \\
& + \frac{ (1-w)\xi^2 }{c\Gamma(a)} \MeijerG*{2}{1}{2}{3}{1,\frac{\xi^2}{c}+1}{a,\frac{\xi^2}{c},0}{ \frac{x^c}{(b A_0\rho)^c}}.
\end{split}
\label{UOWC_cdf}
\end{equation}

\begin{table}[t]
\begin{center}
\setlength{\tabcolsep}{4.5pt}.
\caption{Parameters of the EGG distribution for different bubble levels 
 BL (L/Min) for fresh water and salty water \cite{parameters}}
\begin{tabular}{|c|c|c|c|c|c|c|c}
\hline
Salinity  & BL & $\omega$ & $\lambda$ & $a$ & $b $& $c$ \\
\hline
Salty Water &  4.7   & 0.2064 & 0.3953 & 0.5307 & 1.2154 & 35.7368\\
Salty Water &  7.1   & 0.4344 & 0.4747 & 0.3935 & 1.4506 & 77.0245\\
Salty Water &  16.5   & 0.4951 & 0.1368 & 0.0161 & 3.2033 & 82.1030\\
Fresh Water &  4.7   & 0.2190 & 0.4603 & 1.2526 & 1.1501 & 41.3258 \\
Fresh Water &  7.1   & 0.3489 & 0.4771 & 0.4319 & 1.4531 & 74.3650\\
Fresh Water&  16.5   & 0.5117 & 0.1602 & 0.0075 & 2.9963 & 216.8356\\
\hline
\end{tabular}
  \label{table}
  \end{center}
\end{table} 

\subsection{Overall SNR of dual-hop UAV-assisted 
RF-UOWC system}

As fixed gain AF relaying is adopted for dual-hop UAV-assisted mixed
RF-UOWC system, overall end-to-end SNR at the destination is determined as \cite{snr}
\begin{equation}
\begin{split}
\gamma_{\rm eq} = \frac{\gamma_1 \gamma_2}{\gamma_2 + C}
\end{split}
\label{allSNR}.
\end{equation}

\section{Outage Probability Analysis}

The outage probability refers to the probability  that the instantaneous end-to-end SNR drops below a predefined outage threshold, denoted as $\gamma_{\rm th}$. For the system under investigation, based on the end-to-end SNR at the destination in \eqref{allSNR}, the outage probability can be determined as
\begin{equation}
\begin{split}
P_{\rm out}=\mathrm{Pr}\bigg (\frac{\gamma_{1}\gamma_{2}}{\gamma_{2}+C}<\gamma_{\rm th}\vert \gamma_{2}\bigg)
\end{split}
\label{outage},
\end{equation}
where $\mathrm{Pr}(\cdot)$ denotes probability. Following  some mathematical manipulations, \eqref{outage} can be re-written as

\begin{equation}
\begin{split}
P_{\rm out}&=\int_0^\infty\mathrm{Pr}\bigg (\gamma_{1}<\gamma_{\rm th}+\frac{\gamma_{\rm th}C}{\gamma_{2}}\bigg)f_{\gamma_{2}}(\gamma_{2})\mathrm{d}\gamma_{\gamma_{2}}\\
&=\int_0^\infty F_{\gamma_{1}}\bigg(\gamma_{\rm th}+\frac{\gamma_{\rm th}C}{x}\bigg)f_{\gamma_{2}}(x)\mathrm{d}x.
\end{split}
\label{outage1}
\end{equation}

Substituting \eqref{cdf} and \eqref{UOWC_pdf} into \eqref{outage1},  and after some mathematical derivations presented in Appendix A, the final expression for the outage probability is derived as
\begin{equation}
\begin{split}
&P_{\rm out}  = 1- N\sum_{k=0}^{N-1}\binom{N-1}{k}\frac{(-1)^k }{(k+1)} e^{-\frac{(k+1)\gamma_{\rm th}}{\mu_1}} \\
& \times \Bigg(  w \xi^2 \MeijerG*{3}{0}{1}{3}{\xi^2+1}{1,\xi^2,0}{z_1} \\
&+\frac{(1-w)\xi^2c^{-\frac{1}{2}} }{\Gamma(a)(2\pi)^{\frac{c-1}{2}}} \MeijerG*{2+c}{0}{1}{2+c}{\frac{\xi^2}{c}+1}{a, \frac{\xi^2}{c},\frac{1-1}{c},\frac{2-1}{c},\;...\;,\frac{c-1}{c}}{z_2^c}\! \!\Bigg),
\end{split}
\label{outage2}
\end{equation}
where $z_1=\frac{(k+1)\gamma_{\rm th}C}{\lambda A_0\rho\mu_1}$ and $z_2=\frac{(k+1)\gamma_{\rm th}C}{b A_0\rho \mu_1 c}$.

\section{Numerical Results}
This section presents numerical results obtained using the derived analytical expressions for the outage probability. For the UAV-based RF links, the assumed parameters include a transmitted optical power of $P_t = 100$mW, a noise variance of $\sigma_R^2 = -90$ dBm, and a channel power gain at the reference distance of $G_0 = -30$ dB \cite{sensors}. 
The analysis considers the following values for the UOWC parameters: a conversion efficiency of $\eta=0.8$, a transmitted optical power of $P_{opt} = 100$ mW, and a noise power spectral density of $N_0=10^{-21}$ W/Hz. The specific values used for the UOWC parameters can be found in Table I, as shown in \cite{parameters}.
 \begin{figure}[!t]
\centerline{\includegraphics[width=2.83in]{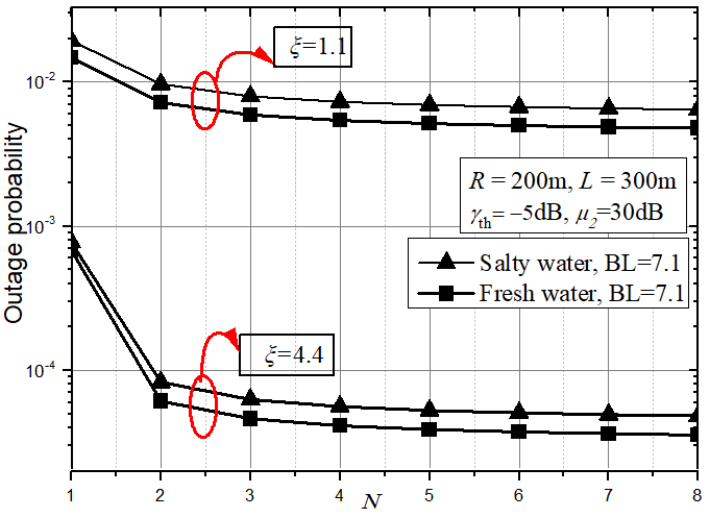}}
\caption{Outage probability vs. number of relays for different values of $\gamma_{th}$, bubble levels and salinity.}
\label{Fig_2}
\end{figure}

Fig.~\ref{Fig_2} depicts dependence between outage probability and number of relays considering: i) different threshold $\gamma_{th}$, ii) different salinity of water, iii) different level of bubbles. Increasing  the threshold SNR, leads to the error probability
deterioration, due to the increase in aggregate interference from other buoys in the system. With greater number of the buoy relays  $N$, the optimal outage probability stays the same, proving that the utilization of greater number of relays up to some optimal point will improve system performance. Beyond the optimal point, further increases in number of relays does not have the impact on the system performance. Analysing number of relays $N$ dependence on level of bubbles $BL$, denser air bubbles leads to performance deterioration due to making turbulence in the water.
In our analysis, we examine both salty and fresh water scenarios. Outage probability is higher for salty water than for the fresh water. This suggests that both water salinity  and lever of air bubbles have impact on system performance, which should be taken into consideration during system design.

\begin{figure}[!t]
\centerline{\includegraphics[width=2.9in]{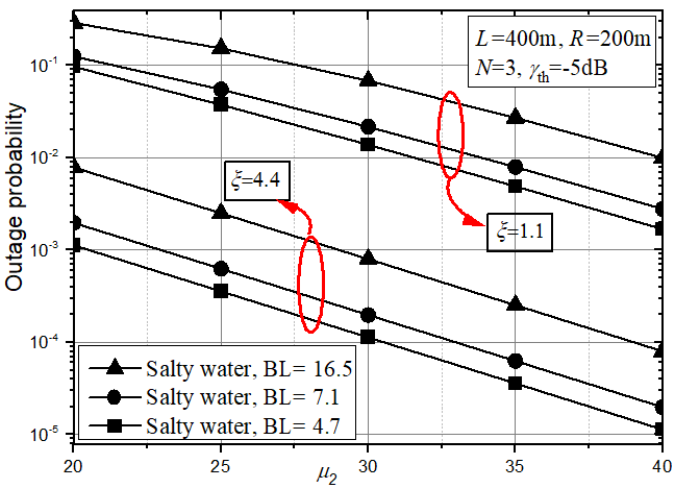}}
\caption{Outage probability vs. average SNR for different values pointing errors  and number of relays.}
\label{Fig_3}
\end{figure}

In Fig.~\ref{Fig_3}, we investigate how pointing errors impact the overall link performance. We analyze this scenario using salty water with a baseline of 16.5 for various numbers of relays. The parameters $A_0$ and $\xi$ dictate the impact of pointing errors. From Figure~\ref{Fig_3}, it's evident that  using parameters $A_0 = 0.5076$ and $\xi = 0.6079$ (indicating weaker pointing errors) yields better results compared to parameters $A_0 = 0.1641$ and $\xi = 0.5244$ (representing stronger pointing errors). As more relays become active in the UOWC system, the outage probability will improve up to some optimal point. 
Once the optimal point is surpassed, additional increases in the number of relays do not significantly impact the system's performance.

\begin{figure}[!t]
\centerline{\includegraphics[width=2.93in]{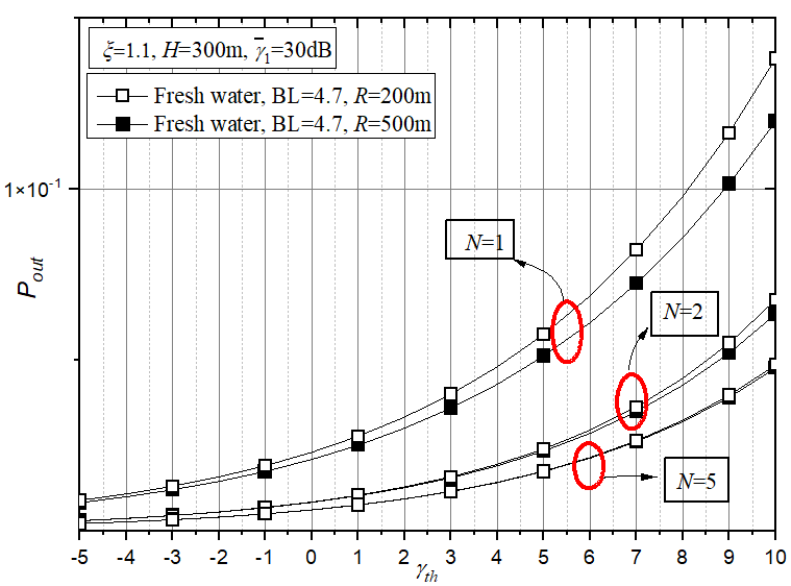}}
\caption{Outage probability vs. radius for different values of $\gamma_{th}$, bubble levels and salinity.}
\label{Fig_4}
\end{figure}

Figure~\ref{Fig_4} illustrates the outage probability dependency on the UAV radius $R$, considering varying numbers of relays and different height of UAV for the circle where buoys are positioned. As expected, when the radius $R$ is larger, increasing $L$ results in performance degradation due to more severe path loss. However, adding more buoy relays improves system performance.  When both $L$ and $R$ are short distances, the overall path loss between buoys and UAV is weak, and the outage probability is lowest. As $L$ increases, distance between buoys and UAV widens, resulting in system deterioration. 

\section{Conclusion}

In this paper, we have analysed  a relay-aided  mixed UOWC-RF downlink system. In the initial phase of the downlink transmission, we encounter the RF communication between hovering UAV and floating buoys. UOWC environment covers second part of the link where, the multiple buoy serve as a relays in communication with underwater devices. Considering the distinct characteristics of both underwater and UAV-based RF channel conditions, we determine the downlink outage probability. 
The numerical results derived from the analysis are utilized to examine the behavior of the outage probability. The system demonstrates improved outage performance under moderate turbulence conditions compared to those under strong turbulence. Moreover, we observe that variations in the number of underwater air bubbles have  impact on system performance. Additionally, it is crucial to consider the positions of UAVs and buoys during the design phase of the system, as these factors  influence overall system performance.

\appendices
\section{}
\label{App1}

The outage probability can be derived by substituting \eqref{cdf} and \eqref{UOWC_pdf} into \eqref{outage1} leading to
\begin{equation}
\begin{split}
&P_{\rm out} = \!\!\int_0^\infty \! \! f_{\gamma_{2}}(x)\mathrm{d}x  \!- \! N\!\! \sum_{k=0}^{N-1} \!\binom{N\!-\!1}{k}\frac{(-1)^ke^{-\frac{(k+1)\gamma_{\rm th}}{\mu_1}} }{(k+1)} \! \\
& \times  \int_0^\infty \!\! e^{-\frac{(k+1)\gamma_{\rm th}C}{\mu_1x}} f_{\gamma_{2}}(x)\mathrm{d}x  \\
& = 1 - N\!\! \sum_{k=0}^{N-1} \!\binom{N\!-\!1}{k}\frac{(-1)^ke^{-\frac{(k+1)\gamma_{\rm th}}{\mu_1}} }{(k+1)} (I_1 + I_2),
\end{split}
\label{outage3a}
\end{equation}
where 
\begin{equation}
\begin{split}
I_1=w\xi^2 \int_0^\infty\!\!\! x^{-1} e^{-\frac{(k+1)\gamma_{\rm th}C}{\mu_1x}} \MeijerG*{2}{0}{1}{2}{\xi^2+1}{1, \xi^2}{ \frac{x}{\lambda A_0\rho}}\mathrm{d}x,
\end{split}
\label{I1}
\end{equation}
\begin{equation}
\begin{split}
I_2 = \frac{(1-w)\xi^2}{\Gamma(a)}\!\!\! \int_0^\infty\!\!\! x^{-1} e^{-\frac{(k+1)\gamma_{\rm th}C}{\mu_1x}}\MeijerG*{2}{0}{1}{2}{\frac{\xi^2}{c}+1}{a, \frac{\xi^2}{c}}{ \frac{x^c}{(b A_0 \rho)^c}}\mathrm{d}x.
\end{split}
\label{I2}
\end{equation}
Integral $I_1$ is solved with the help of  \cite[(01.03.26.0004.01), (07.34.16.0002.01) and (07.34.21.0011.01)]{wolfram} as
\begin{equation}
\begin{split}
I_1 = w\xi^2 \MeijerG*{3}{0}{1}{3}{\xi^2+1}{1,\xi^2,0}{\frac{(k+1)\gamma_{\rm th}C}{\lambda A_0\rho \mu_1}}
\end{split}
\label{I1a}.
\end{equation}Integral $I_2$ can be derived using \cite[(01.03.26.0004.01), (07.34.16.0002.01) and (07.34.21.0013.01)]{wolfram} as
\begin{equation}
\begin{split}
I_2 & =\frac{(1-w)\xi^2 }{\Gamma(a)} \frac{c^{-\frac{1}{2}}}{(2\pi)^{\frac{c-1}{2}}}\\
& \times \MeijerG*{2+c}{0}{1}{2+c}{\frac{\xi^2}{c}+1}{a, \frac{\xi^2}{c},\frac{1-1}{c},\frac{2-1}{c}\;...\;,\frac{c-1}{c}}{\frac{((k+1)\gamma_{\rm th}C)^c}{(b A_0 \rho \mu_1 c)^c}}.
\label{I2a}
\end{split}
\end{equation}
Note that the solution of $I_2$ is valid only if the parameter $c$ is a positive integer. Based on the values presented in Table I, we can round down the value of $c$ from Table I to the nearest lower integer. 
After substituting \eqref{I1a} and  \eqref{I2a} into  \eqref{outage3a}, the final outage probability expression is derived in  \eqref{outage2}.

\vspace{12pt}
\color{red}

\end{document}